# Wavepacket and its collapse

Raoul Nakhmanson[1]

*The notions of wavepacket and collapse are discussed and
a local-realistic interpretation of Berkeley experiment is done.*

## 1. Introduction[2]

On December 14, 1900, at the assembly of German Physical Society in Berlin, Max Planck had presented his theory of radiation [1]. This date is accepted as the birthday of quantum physics, its centennial was marked over the world. In November-December 2000 in Frankfurt-am-Main there was a series of lectures devoted to quantum mechanics. The first lecturer was Carl Friedrich von Weizsäcker, apprentice and colleague of Werner Heisenberg. After the lecture von Weizsäcker answered questions. "How explains physics a connection between matter and consciousness?" - I had asked. "Very good question," – he said, - "but we need two thousand years more to understand this connection."

1900-1902 years were very fruitful also in other respects too. During these three years nine future Nobel prize winners were born, among them Heisenberg (winner 1932), Dirac (1933), Fermi (1938), Pauli (1945), and Wigner (1963). Their personal centennials were also marked. For that reason, in September 2001 in Bamberg (Germany) the Heisenberg Centennial Symposium took place. One of the Bamberg lectures was recently published [2] and was a motive to write this paper.

## 2. Terminology

I will use the notions "wavepacket" and "collapse" in the following widespread sense. Wavepacket is a particular case of a wavefunction. It occupies a restricted region of space and generally moves in space with some velocity. It is to emphasize that the wavefunction is in the *configuration* (*C*-) space. Between the *C*-space (or its subspace) and a real 3D-space can exist a one-to-one correspondence. A wavepacket is difficult to draw and we (as usual) draw only its positive envelope. But we remember enclosed waves which are responsible for interference.

In context with "collapse" we speak about "reduction". Historically the semantics of "reduction" is decreasing and is a simplification. Semantics of "collapse" is a catastrophic worsening of health. Taking it into account it seems meaningful to see "collapse" as a limit case of "reduction", i.e. "reduction up to zero". But commonly both notions are used with the same and even more general meaning like "alteration". It does not lead to confusion because of context.

---

[1] nakhmanson@t-online.de
[2] A variant of Introduction including unauthentic stories can be sent free on request.

## 3. Hole in Heisenberg's logic

I have no possibility to read the original of Heisenberg's text [3], so I reproduce a part of it as cited in [2] with my underlining:

> "In relation to these considerations, one other idealized experiment (due to Einstein) may be considered. We imagine a photon which is represented by a wave packet built up out of Maxwell waves[3]. It will thus have a certain spatial extension and also a certain range of frequency. By reflection at a semi-transparent mirror, it is possible to decompose it into two parts, a reflected and a transmitted packet. There is then a definite probability for finding the photon either in one part or in the other part of the divided wave packet. After a sufficient time the two parts will be separated by any distance desired; now <u>if an experiment yields the result that the photon is, say, in the reflected part of the packet, then the probability of finding the photon in the other part of the packet immediately becomes zero</u>. The experiment at the position of the reflected packet thus exerts a kind of action (reduction of the wave packet) at the distant point occupied by the transmitted packet, and one sees that this action is propagated with a velocity greater than that of light. However, it is also obvious that this kind of action can never be utilized for the transmission of signals so that it is not in conflict with the postulates of the theory of relativity."

One can note following. Because a wavefunction is in *C*-space, its collapse happens there as well. One can agree with Heisenberg that in the situation being under discussion the Schrödinger and the Maxwell equations are formally identical. But the first one works in the *C*-space whereas the second ones are written for the real space. Heisenberg erroneously changes the spaces: From *C*-space which can occupy very small part of real space (see below) he goes to "infinite" real space. This leads him to nonlocality.

## 4. Wavepacket

A photon as a particle has a definite frequency, it is in each instant in a definite point of real space and moves in a definite direction with the light velocity (in vacuum or medium). Such a deterministic conception of a photon was introduced in fundamental works of Planck and Einstein. From such a point of view the sentences like "a photon has a bandwidth" are wrong. The notion "bandwidth" can be applied only to an ensemble of photons.

The frequency of the photon is its immanent inside essence, its pulse, time of its processor. The frequency develops itself outside via energy and interference, and, if there are many photons, via the frequency of electromagnetic field originated.

Quantum mechanics describes the dynamics of a wavefunction. In respect of individual particles it predicts only possibilities, e.g. to have energy $E \pm dE$ to be inside space-time interval $(x_i \pm dx_i, t \pm dt)$. As Einstein said in 5[th] Solvay Congress (1927), the wavefunction describes ensembles rather than individual particles.

---

[3] Heisenberg added the following footnote here: "For a single photon the configuration space has only three dimensions; the Schrödinger equation of a photon can thus be regarded as formally identical with the Maxwell equations."

The wave function of an ensemble of identical photons is pure sinusoid. It means the photon can be found with the same possibility in any point of the huge *real* space. This non-practical idealization appears because we ignore history. We don't have an abstract photon, it was once emitted by an atom.

During a transition between two states the atom as a real physical system can emit a photon in some frequency range. The interference of these *possible* frequencies builds wavepacket in *C*-space. The projection of the wavepacket square module on the real space shows where, when, and with which possibility the photon can be found. The black curve in Fig.1 gives an example of such a packet. It moves from the atom with light velocity and keeps its form if the medium has no dispersion.

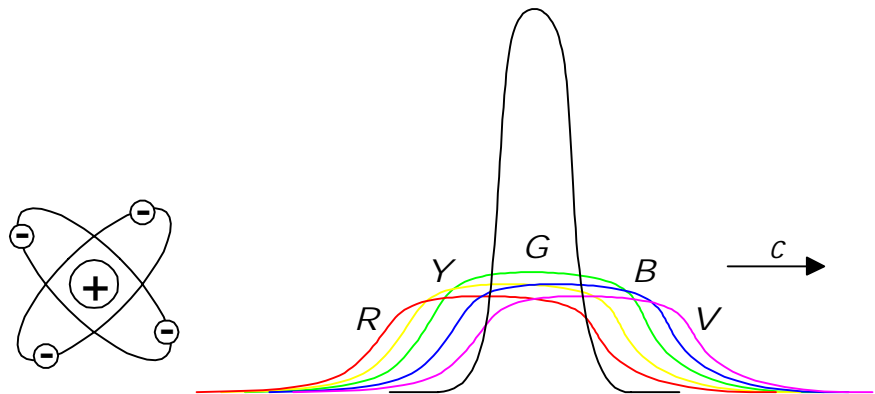

Fig.1. Wavepacket emitted by the atom and his color subpackets.

The "discharge" of a photon from an atom opposes some energetic barrier. Photons with higher energy discharge earlier, so the possibility to find a high-frequency photon is higher in the front of the packet. Therefore the total wavepacket can be decomposed into particular subpackets differing from each other not only by color but also by position (Fig.1, the colors are marked conditionally: *R*-red, *Y*-yellow, *G*-green, *B*-blue, *V*-violet). If one reduces the frequency bandwidths of sub-packets, the number of sub-packets increases. Simultaneously increase the packet widths (i.e. coherence lengths) of sub-packets. The width of the total wavepacket is defined by the atom and stays constant.

As said, the wavepacket is in *C*-space. Its projection on the real 3D space can be a real material packet if many photons would be emitted by many identical atoms. Practically it means that we are dealing with a light impulse containing thousands of photons. They form a classical object, that is, an electromagnetic field. Atomism (indivisibility, individual) vanishes: If such a packet meets a semitransparent mirror it divides in two packets similar to the original one. The photon is not built from Maxwellian (classical) waves, on the contrary, Maxwellian waves are built from photons (quantum field theory). Transition from quanta to classic is a transition from individual to crowd, it is transition from quantity to quality.

If a such real packet corresponding to Fig.1 meets an energy barrier the probability to overcome it is for "violet" photons greater than for "red" or "green" ones. The original packet divides in transmission one, e.g. *V*, and reflection one e.g. *R+Y+G+B*. Because the maximum of "violet" packet *V* forestalls the maximum of original packet *R+Y+G+B+V*, an observer can conclude that the packet *V* passes the barrier region with superluminal velocity. Such an "acceleration" was really seen in [4].

## 5. Collapse

Where is the *C*-space containing wavefunction with its collapse? At the beginning of this paper I remember the series of lectures in Frankfurt 2000. One of the lecturers was A.Zeilinger from Vienna. The morning after the lecture he answered questions, and I had asked him about it: His answer was: "In my head!". The idea isn't new, 70 years ago it was discussed by J.von Neumann, E.Wigner, F.London and E.Bauer. Their opinion is shared up to now by some people. Of course, there are some connections between mind and matter. Human consciousness can receive information coming from organs of sense, and works it on. The consciousness can control nerve impulses sending to muscles, and we can speak, write, and stay things on the path of light beam. Perhaps we need really 2000 years to understand how this machinery works. But it works *inside* us! Perhaps in some special cases (parapsychological phenomena if they exist?) it works also outside. If so, these cases are exceptions: In the ten billion years after Big Bang particles have been moving obeying the same rules as today even though there were no people. Therefore we must seek the wavefunction in the particle, more precisely, in its consciousness [5]. In human consciousness can be only some idea about this function, right or not.

In human consciousness is its own "wavefunction" and its *C*-space[4]. It is a strategy of the person, its plans for the future. If there are several alternative ways to the goal they interfere in the consciousness of the person. The laws of forming of this wavefunction are based on some general principle ("principle of least action"?) and can be found studying the human behavior. The wavefunction gives only "weights" of alternatives. The choice is not definitely in favor of a heavy alternative but would be randomized taking into account the weights. Such a tactics allows to search all possibility. Therefore the laws of psychology and sociology (present and/or future) - like to quantum-mechanical laws - can "exactly" describe only ensembles. In respect of individuals they can predict only possibilities.

The indivisible individual chooses one alternative and excludes the rest. This means a reduction/collapse of it's wave function. It occurs in the human consciousness and controls events in real space. One has information only about the past. To plan one's behavior successfully one must predict the future. If the prediction is deep enough the choice can be done very before it is non-reversible physically. Receiving of new information opens new alternatives. Therefore the collapse of the wavefunction is a reiterative event in the *C*-space.

Two or more persons can have common plans i.e. a common wave function. Being separated in space they continue to act in concord up to deepest level of their prediction of future being at the moment of their separation. An example of a common (entangled) wavefunction is in [6].

The hypothesis that elementary particles and atoms like people and other biological objects have some kind of consciousness allows naturally explain apparent paradoxes of the quantum world and suggest experiments to check itself [5]. This hypothesis could be proposed in natural science as far back as 19th century with experimental justifications of atomism: Known indivisible (in system meaning, e.g. atom) objects, including people and products of their labor, have within its species very similar physical characteristics and possess some consciousness or are made with its help.

---

[4] Don't confuse with quantum-mechanical wavefunction which can be formally written for a cat, a man, or for whole universe ...

To the situation discussed by Heisenberg and our critique given in Sec.3 one can add that Heisenberg's sentence

> "if an experiment yields the result that the photon is, say, in the reflected part of the packet, then the probability of finding the photon in the other part of the packet immediately becomes zero" is not correct: the photon is in real space whereas wavepacket is in $C$-space. The collapse of the wavepacket i.e. choice of reflection from a semitransparent mirror occurs in the consciousness of the photon much earlier than when the photon was found in the real space. The collapse occurs not later than the moment of interaction between photon and mirror, and perhaps even when the photon detached itself from the atom. The registration of a photon means only a collapse of knowledge in the consciousness of an experimenter.

## 6. Berkeley experiment

In the paper [2] an experiment performed by authors in Berkeley is described. Fig.2 shows its schema with some simplifications which do not matter to the principle.

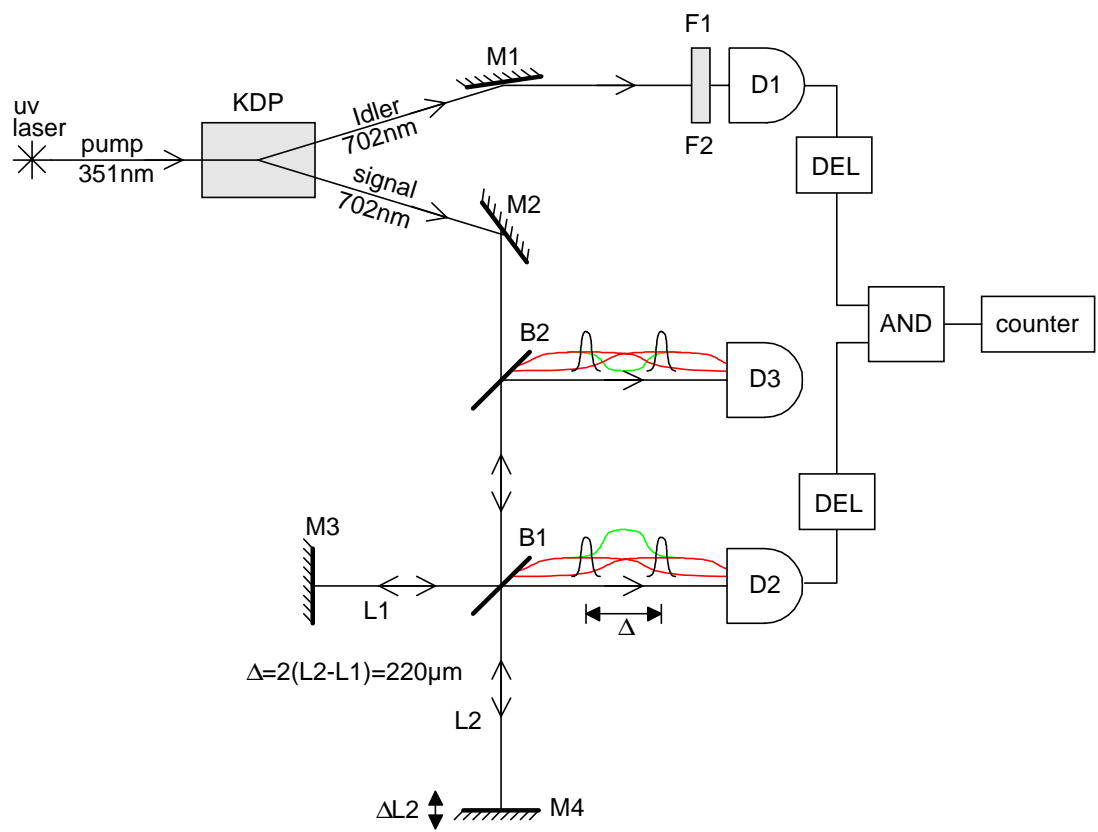

Fig.2. Berkeley experiment. M are the mirrors, B are the beamsplitters, F are the filters, D are the detectors, DEL are delay lines, AND is the coincident unit.

"Pump" photons generated by the argon laser with the wavelength of 351 nm penetrate in KDP crystal and split in signal-idler pair. The conservation of energy and momentum is fulfilled. Because the laser beam is very monochrome the sum of energies of signal and idler photons (but not their individual energies) is sharp defined. After the KDP the screen with two pinholes (not shown in Fig.2) selects for both signal and idler ensembles a middle wavelength λ ≈ 702 nm and bandwidth Δλ ≈ 25 nm. The idler photons are registered by the "removed" detector D1. Before it one can place a narrowband filter F1 with bandwidth Δλ = 0.86 nm centered at λ = 702 nm or a broadband filter F2 with bandwidth Δλ = 10 nm centered at λ = 702 nm. The signal photons entered a Michelson interferometer (beamsplitter B1 and mirrors M3 and M4) and were registered by the detector D2. The beamsplitter B2 and the detector D3 in Fig.2 are placed differently from an original Berkeley schema. These elements didn't play a principal role. In Fig.2 they remember that the interferometer has two outputs.

Signals from D1 and D2 after delay lines DEL came to a coincidence unit AND with time window 1.0 ns and were stored into a counter. The noise level and intencity of light were low, so a main part of coincidences was indebted to signal-idler pairs. The arms of Michelson interferometer L1 and L2 have different lengths, the optical path difference Δ =2(L2-L1) = 220 µm. The length L2 can be smoothly changed to register an interference.[5]

R.Chiao and P.Kwiat, the authors of the paper [2], investigated the count rate vs. the length L2. If before D1 was no filter or the broadband filter F2 then the change of L2 in ≈1 µm (change of optical path Δ in ≈2µm) the count rate didn't practically change. But if the narrowband filter F1 was before D1 then the periodical change of count rate corresponding to the interference of waves with λ = 702 nm was observed.

For the wavelength λ and bandwidth Δλ the coherence length i.e. wavepacket width W is

$$W = \lambda^2/\Delta\lambda \ .$$

For λ = 702 nm and Δλ=25, 10, or 0.86 nm wavepacket widths equal 20, 50, and 570 µm, respectively. In two former cases it is shorter and in the last case more than the interferometer optical path difference Δ=220 µm.

The authors of [2] explain their experimental results as following: Because the signal and idler photons are in an entangled state of energy, introducing of the "removed" filters F1 and F2 causes an instant nonlocal collapse of a wavefunction of the signal ensemble decreasing its bandwidth to the bandwidth of the idler ensemble and, respectively, increasing its wavepacket width. For filter F2 W=50 µm < Δ=220 µm therefore the wavepackets passing different interferometer arms don't overlap each other and we don't see interference. For filter F1 W=570 µm > Δ=220 µm, the wavepackets overlap each other and interfere. The interference depends on a phase difference hence on Δ. Below are four citations from [2] (*italic* as in the original, underlining mine):

---

[5] In [2] not geometrical but optical length of L2 was changed, but it is not a matter of principle.

We shall see that a measurement of the energy of one daughter photon has an instantaneous collapse-like action-at-distance upon the behavior of the other daughter photon. (end of Sec.1 of [2])

In order to conserve total energy, the energy bandwidth of the collapsed signal photon wavepacket must depend on the bandwidth of the filter F1 in front of D1, through which it did not pass. Therefore, the visibility of the signal photon fringes seen in coincidences should depend critically on the bandwidth of this *remote* filter. For a narrow-band F1, this fringe visibility should be high, provided that the optical path length difference of the Michelson does not exceed the coherence length of the *collapsed* wavepacket (recall that due to the energy-time uncertainty principle, collapsing to a *narrower* energy spread actually leads to *longer* wavepacket). It should be emphasized that the width of the collapsed signal photon wavepacket is therefore determined by the *remote* filter F1, *through which this signal photon has apparently never passed!* If, however, a sufficiently broadband remote filter F1 is used instead, such that the optical path length difference of the Michelson is much greater than the coherence length of the collapsed wavepacket, then the coincidence fringes should disappear. (end of Sec.4 of [2])

The observed visibility of the coincidence fringes was quite high, viz., 60% ± 5%, indicating that the collapse of the signal photon wavepacket had indeed occurred. (beginning of Sec.5 of [2])

In conclusion, we have demonstrated that the *nonlocal* collapse of the wavefunction or wavepacket in the Copenhagen interpretation of quantum theory, which was introduced by Heisenberg in 1929, leads to a self-consistent description of our experimental results. Whether or not fringes in coincidence detection show up in a Michelson interferometer on the near side of the apparatus, depends on the arbitrary choice by the experimenter of the remote filter F1 through which the photon on the near side has evidently never passed. This collapse phenomenon, however, is clearly *noncausal*, as a "delayed-choice" extension of our experiment would show. (end of Sec.8 of [2])

Unfortunately I must disillusion the authors. In their experiment there are no collapse of the signal wavepacket indebted to "remote" filters F1 and F2. The same broadband ensemble of signal photons reached detectors D2 and D3. To make it certain the authors can measure directly a spectrum of signal photons. It follows also from a fact noticed by the authors that the intensity of light measured by detectors D2 and D3 directly (i.e. without coincidence unit) does not depend on filters F1 and F2. Besides, the collapse introduced by the authors really allows superluminal communication. To have it one must cancel D1, D3, AND, and interferometer. The signal photons must be directed simply to the detector D1. An intensity of the pump laser can be increased and the bandwidth of F1 can be decreased, better to zero. Now, introducing and removing F1 in/from the "remote" idler beam one can instantaneously (i.e. without transmission delay) modulate a number of photons registered by the D2. To increase information flow an electric modulator is favorable. But I must warn the readers planning to build such a superfast telegraph: It would not work at all.

An explanation of the Berkeley-experiment results lies in peculiarities of the coincidence circuit. Introducing of filters F1 or F2 reduced only the spectrum of idler ensemble registered by the detector D1. The detectors D2 and D3 registered as before the whole signal ensemble with the spectrum cut out by the screen with pinholes, i.e.

with Δλ=25 nm. But the coincidence circuit "saw" only signal photons belonging to a subensemble which is complementary to idler ensemble registered by the detector D1. Signal photons belonging not to this subensemble the coincidence circuit ruled out. Because of energy correlation this signal subensemble had exactly the same bandwidth as the idler ensemble cut out by the filters F1 and F2. Because of linearity and superposition principle we can consider this subensemble independently.

Fig.2 shows the wavepackets of signal subensembles so as they are "seen" by the coincidence unit.[6] The space interval between packet centers is equal to the interferometer arm length difference Δ=220 µm. With a black color are shown the subpackets selected when the broadband filter F2 with bandwidth Δλ=10 nm is placed before the detector D1. As explained, the same bandwidth must have the complementary signal subpackets. The consequent wavepacket widths W are 50 µm that is less as Δ. Therefore these subpackets have no common space-time interval and don't interfere. With a red color are shown the subpackets selected when the narrowband filter F1 with bandwidth Δλ=0.86 nm is placed before the detector D1. The consequent wavepacket widths W are 570 µm that is more than Δ. Therefore these packets have a common space-time interval and interfere. This interference depends on the phase difference which in its turn depends on the Δ. The red curves in Fig.2 don't take interference into account. The green curves show the result of interference for a case when in direction to D2 it is constructive and in direction to D3 it is destructive. Therefore the alteration of Δ changes interference and the number of signal photons registered by D2 as it was found in [2]. Visibility of interference V depends on W and Δ and equal

$$V = 1 - \Delta/W \quad (0 \leq \Delta \leq W) \quad .$$

Substitution of Δ=220 µm and W=570 µm (filter F1) leads to V=0.614 i.e. 61.4% that is in a good accordance with experimental value V=60±5% found in [2].

## 7. Conclusion

In 1835 the story "The Emperor's New Clothes" was published in Copenhagen. In this story thanks to a little boy the common sense took the upper hand. One hundred years later in the same Copenhagen the Queen of Physics – quantum mechanics – also became new clothes, so called "Copenhagen interpretation". Yet then these clothes seemed very ephemeral. "Bad boys" like Einstein cried but were not heard. Time makes these clothes more and more ephemeral. Why not try the new clothes "made in Frankfurt"? Up to now nobody finds holes in it. A.Zeilinger in Frankfurt said categorically: "Your theory cannot be refuted!" Strangely but it is regarded as fatal disadvantage. Here is an essence from e-mail correspondence with A.Elitzur:

A.E.: "The only question is whether this explanation is scientific. Here our strong disagreements remained. I still think, however, that your approach deserves to be considered among other prevailing interpretations of quantum mechanics."
R.N.: "Please let me know your definition for 'scientific'."

---

[6] The original wavepackets are, of course, in the *C*-space.

A.E.: "I believe in Popper's criterion. If you can point out an experiment that can definitely refute your interpretation then I would call it scientific. I guess with some effort you can conceive of such an experiment and it will be worth performing."

R.N.: "I have read Popper. I can not remember that he defines 'scientific' as you do it."

So far as I remember Popper the Elitzur logic seems as following:
    a) "A theory cannot be proved scientifically, it can be only refuted" (Popper).
    b) "I cannot refute this theory and the author doesn't help me" (Elitzur).
    c) "Therefore the theory is not scientific" (Elitzur).

Albeit I suggest experiments to confirm my hypothesis and to go beyond of border of quantum mechanics it is not of interest and is not scientific. ☹

    Practice shows that chemistry can be reduced to physics. Neurophysiologists hope that the mental processes can be explained with use of quantum mechanics i.e. also reduced to physics. But physicists, among them Einstein and Bohr, did not share this idea. In the light of the "intelligent particle" hypothesis it is naturally to suggest that physics itself can be reduced to psychology and sociology.